\documentclass{mem}
\usepackage{natbib}\usepackage{txfonts}\usepackage{balance}
\usepackage{graphicx}
\usepackage[a4paper]{hyperref}
\begin{document}
\def\teff{$T\rm_{eff }$}
\def\kms{$\mathrm {km\,s}^{-1}$}
\def\ms{$\mathrm {m\,s}^{-1}$}

\title{
Chromospheric heating and structure as determined from high resolution 3D simulations
}

   \subtitle{}

\author{
M.\,Carlsson\inst{1,2} 
\and V.H.\,Hansteen\inst{1,2}
\and B.V.\,Gudiksen\inst{1,2}
          }

  \offprints{M.\,Carlsson}
 
\institute{
Institute of Theoretical Astrophysics, University of Oslo, P.O. Box 1029 Blindern, N-0315 Oslo, Norway
\and
Center of Mathematics for Applications, University of Oslo, P.O. Box 1052 Blindern, N-0316 Oslo, Norway
}

\authorrunning{Carlsson}

\titlerunning{Chromospheric 3D simulations}

\abstract{ We have performed 3D radiation MHD simulations extending
  from the convection zone to the corona covering a box 16 Mm$^3$ at
  32 km spatial resolution. The simulations show very fine structure
  in the chromosphere with acoustic shocks interacting with the
  magnetic field. Magnetic flux concentrations have a temperature
  lower than the surroundings in the photosphere but higher in the low
  chromosphere. The heating is there mostly through ohmic dissipation
  preferentially at the edges of
  the flux concentrations. The magnetic field is often wound up around
  the flux concentrations. When acoustic waves travel up along the
  field this topology leads to swirling motions seen in chromospheric
  diagnostic lines such as the calcium infrared triplet.
   }
\maketitle{}

\section{Introduction}\label{sec:introduction}

Inferring physical conditions in the solar chromosphere is difficult,
because the few spectral lines that carry information from this 
region (e.g.\ hydrogen Balmer-$\alpha$, Ca\,{\sc II} H and K lines
and infrared triplet and He 1083~nm line) are formed outside
local thermodynamic equilibrium (LTE). Thus they do not directly couple
to the local conditions. In addition, the chromosphere is very
dynamic with large variations in temperature and density and 
a concept like ``formation height'' is not very useful. We also
go from plasma domination in the photosphere to a magnetically
dominated regime in the upper chromosphere. This means that
the chromosphere can sustain a multitude of wavemodes that
become degenerate and couple when the sound speed equals 
the Alfv\'en speed. Major progress in our understanding is 
dependent on comparison between detailed observations and
detailed modelling. Such modelling needs to go all the way from the
convection zone (where waves are excited and magnetic field foot-point braiding
takes place) to the corona (to enable magnetic field connectivity) and
include a reasonable treatment of the radiative transfer.
The first such modelling was reported by \citet{Hansteen2004}. These
simulations have been used to study flux emergence through the
photosphere into the corona
\citep{Martinez-Sykora+Hansteen+Carlsson2008,
  Martinez-Sykora+Hansteen+Carlsson2009} and driving of spicules
\citep{Martinez-Sykora+Hansteen+DePontieu+Carlsson2009}. 
We here analyze a high resolution simulation focusing on the
chromospheric structure, heating and dynamics close to magnetic flux
concentrations. 
The layout of this contribution is as follows: in
Section~\ref{sec:methods} we outline the methods used, in
Section~\ref{sec:setup} we describe how the simulation analyzed was
set up, in
Section~\ref{sec:results} we give the results and we finish in
Section~\ref{sec:discussion} with conclusions.

\section{Methods}\label{sec:methods}

Earlier simulations were carried out with the ``Oslo Stagger
Code'', a code where the parallelization was performed assuming shared memory
and therefore restricted to rather few CPUs, see
\citet*{Hansteen+Carlsson+Gudiksen2007} for a description. We have now
completed a full rewrite using the Message Passing Interface (MPI)
parallelization library and a completely new radiation transfer module
employing domain decomposition
\citep{Hayek+Asplund+Carlsson+Gudiksen+etal2010}. This new code, named
BIFROST, is outlined below. A detailed writeup is in preparation
\citep{BIFROST}.

We solve the equations of radiation-magnetohydrodynamics using a
compact, high-order, finite difference scheme on a staggered mesh
(6th order derivatives, 5th order interpolation). The equations are 
stepped forward in time using the explicit 3rd order
predictor-corrector procedure of \citet{Hyman1979}, modified for
variable time steps. High-order artificial diffusion is added both
in the forms of a viscosity and in the form of a magnetic diffusivity.
The radiative transfer equation is solved along 24 rays using a
domain-decomposed short-characteristic method. The opacity from millions of
spectral lines is included using multi-group opacities in four bins 
\citep{Nordlund1982} modified to include the effects of coherent
scattering \citep{Skartlien2000}. Conduction along the magnetic
field-lines is solved implicitly using a multigrid method. In addition
to the radiative energy exchange treated with the detailed solution
of the radiative transfer (including scattering effects) we include
optically thin radiative losses in the the corona and radiative losses
from strong lines of hydrogen and singly ionized calcium. These latter
losses dominate the radiative losses from the chromosphere
\citep{Vernazza+Avrett+Loeser1981}. We include these radiative losses
without assuming LTE by using an expression

$$\Phi=\theta(T)e^{-\tau}f(\tau)$$

\noindent where $\Phi$ gives the radiative losses per electron per atom from a given element (hydrogen and
singly ionized calcium, respectively), $\theta(T)$ is the sum of all
collisional excitations (obtained from atomic data), $T$ is
temperature, $\tau$ is an optical depth and $e^{-\tau}f(\tau)$ is an
effective escape probability. The optical depth $\tau$ is set 
proportional to the vertical column mass with the proportionality
constant obtained from the 1D detailed non-LTE radiation hydrodynamic
simulations of Carlsson \& Stein (1992, 1995, 1997, 2002)
\nocite{Carlsson+Stein1992}
\nocite{Carlsson+Stein1995}
\nocite{Carlsson+Stein1997}
\nocite{Carlsson+Stein2002}
The function $f(\tau)$ is obtained from a fit using the same 1D
simulations.

\section{Simulation setup}\label{sec:setup}

The simulation analyzed in this contribution covers a region
16.6$\times$16.6$\times$15.5 Mm$^3$ with the bottom boundary in the
convection zone 1.4 Mm below the height where the mean optical depth at 500~nm is
unity (in the following this height defines our z=0) and the top
boundary in the corona 14.1 Mm above this height. Both the bottom and
top boundaries are transparent. The simulation domain is periodic
in both the horizontal directions with a fixed spacing of
32.5 km while the vertical grid size is 24-28 km up to 4 Mm
height, increasing gradually above this height to a final maximum
value of 213 km at the top of the computational domain. The final
simulation has 512$\times$512$\times$325 grid-points. The initial
relaxation of the convection was performed at lower resolution.

The mean, unsigned field strength at z=0 is initially 40 G but falls to about
10~G at the beginning of the high-resolution run due to the fact that no 
field is injected into the computational domain and the existing, tangled field,
decays slowly. 

\section{Results}\label{sec:results}

\begin{figure}[htb!]
\resizebox{\hsize}{!}{\includegraphics[clip=true]{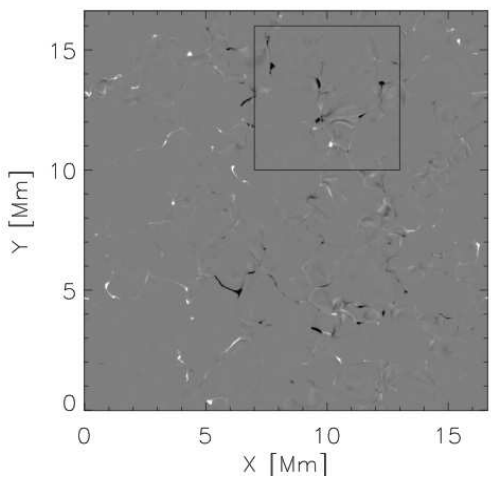}}
\caption{\footnotesize
Vertical magnetic field strength at z=0, t=1500\,s. The colour scale has been
clipped at 500 G to emphasize weaker field. The maximum field strength
is $\pm$ 1~kG. The square marks the region of interest shown in more detail
in Fig.~\ref{fig:tbq}.
}
\label{fig:bz_150}
\end{figure}

After the increase of the resolution, there is a slow increase in the
spatial power at high frequencies. After some 400\,s the evolution
starts to diverge from the continuation of the low resolution run ---
granules fragment more easily and the velocity power at small spatial
scales increases. The mean unsigned field strength also starts to
increase, presumably due to local dynamo action made possible at the
higher resolution, and the mean unsigned field strength ends up at 15
G around t=1500\,s and stays at this level until the end of the
simulation at t=1670\,s. 
%
%
The vertical magnetic field strength at z=0,
t=1500\,s is shown in Fig.~\ref{fig:bz_150}. As can be expected from
the low average unsigned field of 15~G there are large regions with
only weak field. There are several regions with one polarity
dominating but in general, both polarities co-exist. This is even more
evident if the image scaling is adjusted to emphasize the weak field.
The magnetic flux concentrations have a field strength close to 1~kG.

\begin{figure}[htb!]
\resizebox{\hsize}{!}{\includegraphics[clip=true]{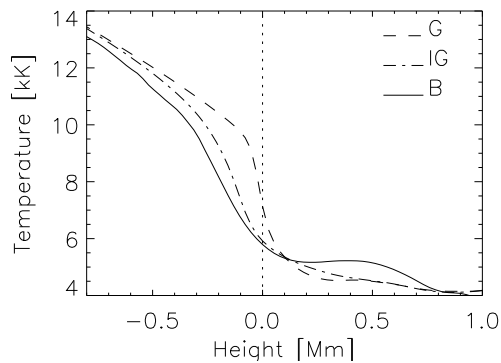}}
\caption{\footnotesize
Mean temperature as function of height for granules ({\it dashed}), intergranular
lanes ({\it dash-dotted}) and magnetic flux concentrations ({\it solid}).
}
\label{fig:tz_bp}
\end{figure}

The temperature structure of these magnetic flux concentrations
compared with the temperature structure of granules and intergranular
lanes is shown in Fig.~\ref{fig:tz_bp}. The average temperature for
granules has been taken over all columns with a vertical upflow
larger than 100 \ms\ at z=0. The intergranular lane average has been
taken over all columns with a vertical downflow larger than 100 \ms\
and the average for magnetic flux concentrations has been taken over
all columns where the unsigned magnetic field in the photosphere is
larger than 300~G. Intergranular lanes are hotter than granules above
the surface (inverse granulation). Magnetic flux concentrations are
cooler than intergranular lanes below the surface at a given geometric
height but hotter than both granules and intergranular lanes in the
upper photosphere and lower chromosphere.

\begin{figure*}[htb!]
\resizebox{\hsize}{!}{\includegraphics[clip=true]{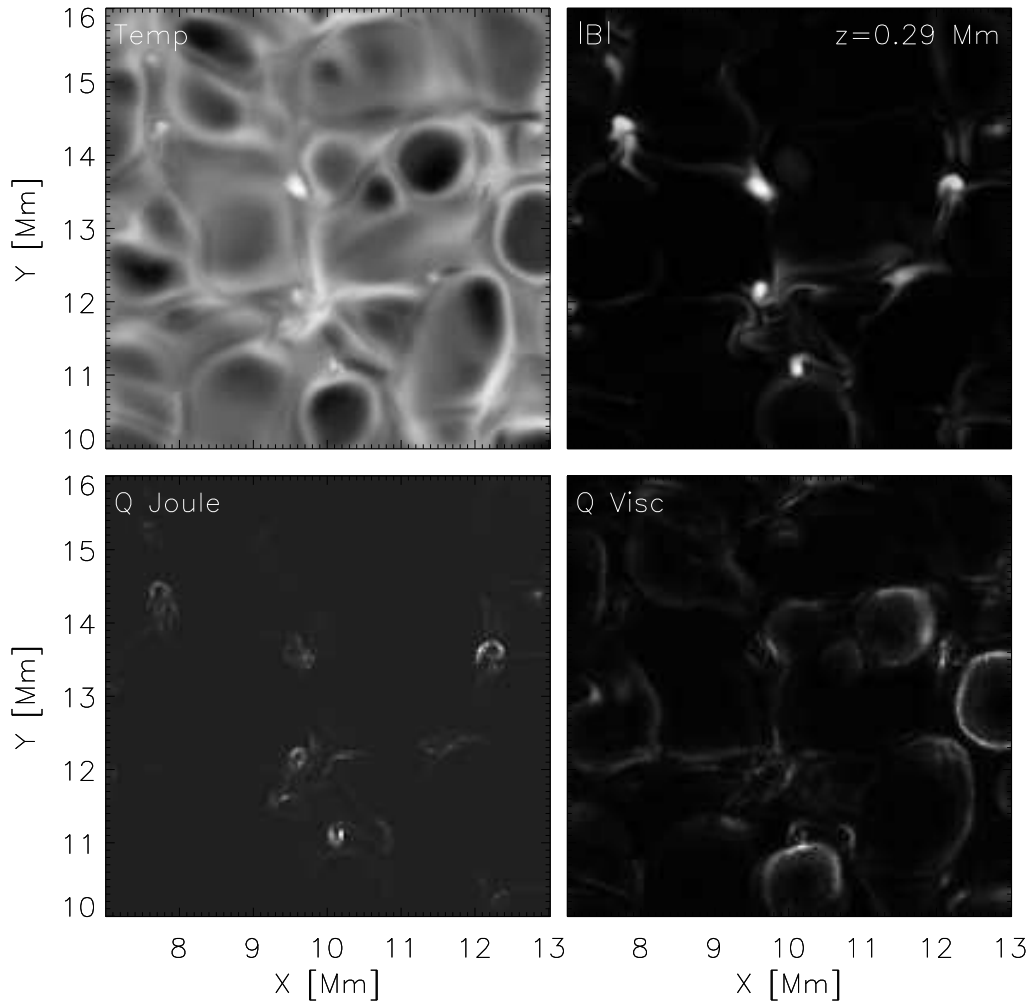}}
\caption{\footnotesize
Temperature ({\it upper left panel}), magnetic field strength ({\it upper right}), 
Joule heating ({\it lower left}) and viscous heating ({\it lower right}) at a height
of 0.29 Mm at t=1500\,s. At this height, magnetic flux concentrations are hotter than
the surroundings due to Joule heating
}
\label{fig:tbq}
\end{figure*}

This increased temperature in the upper photosphere-lower chromosphere of magnetic flux
concentrations is clearly shown in Fig.~\ref{fig:tbq}. The region
shown with a square in Fig.~\ref{fig:bz_150} is shown in more detail
at a height of 0.29 Mm with temperature, magnetic field strength,
Joule heating and viscous heating displayed. It is clear that most
magnetic flux concentrations have a higher temperature than the
surroundings at this height and that Joule heating takes place at the
locations of high magnetic field strength. These locations evolve only
slowly in time while the viscous heating is concentrated at the edges
of granules evolving on much shorter timescales.

\begin{figure*}[htb!]
\resizebox{\hsize}{!}{\includegraphics[clip=true]{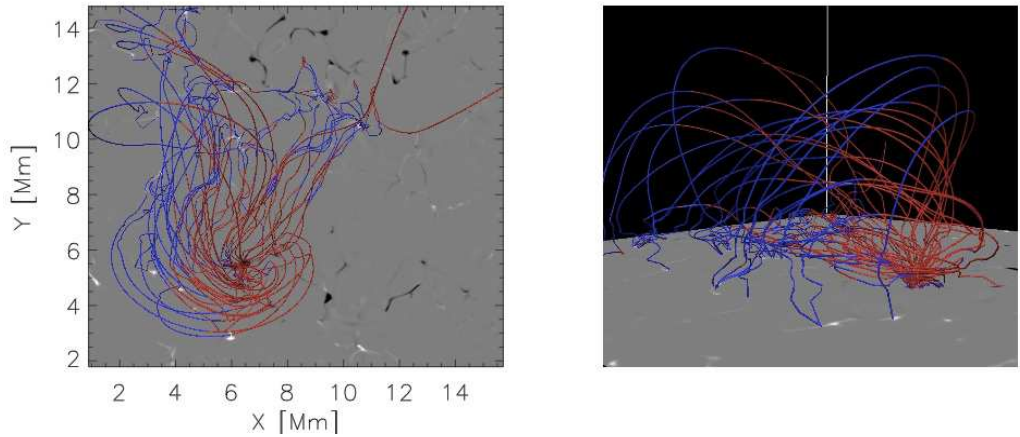}}
\caption{\footnotesize
Magnetic field above a magnetic field concentration as seen from above ({\it left panel}) and
in perspective from the lower left corner ({\it right panel}). The vertical component of the
field at z=0 is shown in greyscale. The sign of Bz is shown
in colour with red field lines corresponding to negative Bz (dark in the plane at z=0). The
highest field-lines shown extend to a height of 5.8 Mm. Note the spiral topology as 
seen from above and the abundance of low-lying almost horizontal field-lines.
}
\label{fig:vapor_field_101}
\end{figure*}

\begin{figure*}[htb!]
\resizebox{\hsize}{!}{\includegraphics[clip=true]{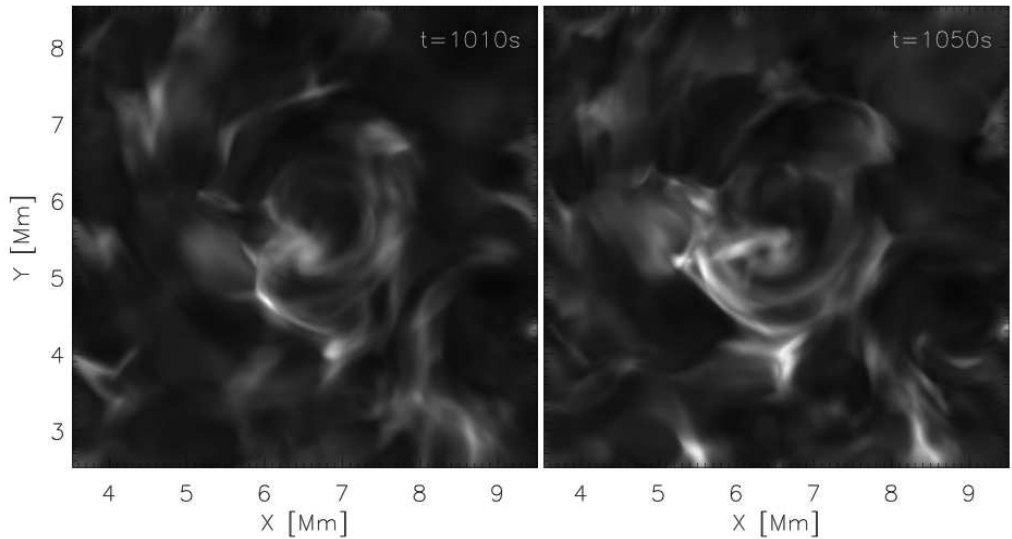}}
\caption{\footnotesize
Synthesized narrow band image at the core of the chromospheric
$\lambda$854.2~nm line from singly ionized calcium at two times
showing the apparent swirling motion.
}
\label{fig:ca8542}
\end{figure*}

Due to foot-point motions, the magnetic field topology in the
chromosphere changes all the time. In the upper photosphere and lower
chromosphere the plasma $\beta$ (ratio between gas pressure and
magnetic pressure) is larger than unity and the magnetic field is far
from force-free. In the upper chromosphere, $\beta$ drops below unity
and the field approaches a force-free state. This interplay between
field-braiding and expansion into a force-free state often leads to
spiral topologies around small magnetic flux concentrations. One
example from the simulation is shown in Fig.~\ref{fig:vapor_field_101}
where the field-topology is shown around a magnetic flux concentration
at t=1010\,s. Field-lines from the negative polarity in the flux
concentration connect to various positive polarity patches with a
clear spiral pattern as seen from above (left panel in
Fig~\ref{fig:vapor_field_101}) and a large amount of almost horizontal
(but curved) field (right panel in Fig~\ref{fig:vapor_field_101}). At this instant in time in the simulation
there is an acoustic wave that turn into a slow-mode wave in the
low-$\beta$ region travelling along the curved field-lines. In
synthesized, narrow band, images from the core of the 
$\lambda$854.2~nm line of singly ionized calcium (Fig.\ref{fig:ca8542}) this is visible as a swirling
motion around the flux-concentration very similar to observations
recently reported by \citet{Wedemeyer-Bohm+Rouppe-van-der-Voort2009}.

\section{Discussion and conclusions}\label{sec:discussion}

A multitude of phenomena are seen in these high resolution
radiation-MHD simulations
that extend from the convection zone to the corona and we have here only
focused on the heating and dynamics around magnetic flux
concentrations. Although these simulations are the most ambitious
performed to date, there are still important ingredients missing. The
current simulations do not include the effects of out-of-equilibrium
hydrogen ionization even though it is known that such effects are very
important for the energy balance of especially the middle-higher chromosphere
\citep{Carlsson+Stein2002, Leenaarts+Wedemeyer-Bohm2006,
  Leenaarts+Carlsson+Hansteen+Rutten2007}. Including such effects
would likely increase the amplitude of shocks in the middle
chromosphere. Another limitation is the neglect of the back-radiation
from the corona onto the upper chromosphere. This radiation would
heat the chromosphere above about 1~Mm height
\citep{Carlsson+Stein2002} but have little effect on the heights
discussed in this contribution. 

\begin{acknowledgements}
This research was supported by the Research Council of Norway through
the grant ``Solar Atmospheric Modelling'' and 
through grants of computing time from the Programme for Supercomputing.
\end{acknowledgements}

\bibliographystyle{aa}

\end{document}